\documentclass[aps,12pt,floatfix,tightenlines,showkeys,showpacs,preprint,nofootnotebib]{revtex4}
\usepackage{amsmath}
\usepackage{amsfonts}
\usepackage{commath}
\usepackage{graphicx}
\usepackage{caption}
\usepackage{subcaption}
\usepackage{float}
\usepackage{epstopdf}
\newcommand{\bfr}{{\bf r}}

\newcommand{\ben}{\begin{displaymath}}
\newcommand{\een}{\end{displaymath}}
\newcommand{\be}{\begin{equation}}
\newcommand{\ee}{\end{equation}}
\newcommand{\bea}{\begin{eqnarray}}
\newcommand{\eea}{\end{eqnarray}}

\newcommand{\bfb}{{\bf b}} 

\newcommand{\bfk}{{\bf k}}

\newcommand{\bfR}{{\bf R}} 
\let\oldhat\hat
\renewcommand{\hat}[1]{\oldhat{\mathbf{#1}}}
\begin{document}
\preprint{NT@UW-14-12}
\title{Corrections to Eikonal Approximation for Nuclear Scattering at Medium Energies}
\author{Micah Buuck and Gerald A. Miller}
\affiliation{Department of Physics,
University of Washington, Seattle, WA 98195-1560}
\date{\today}

\begin{abstract}
The upcoming Facility for Rare Isotope Beams (FRIB) at the National
Superconducting Cyclotron Laboratory (NSCL) at Michigan State
University has reemphasized the importance of accurate modeling of low
energy nucleus-nucleus scattering. 
Such calculations have  been simplified by using the 
eikonal approximation.  As a high
energy approximation, however, its accuracy suffers for the medium energy  beams
that are of current experimental interest. A prescription developed by Wallace~\cite{Wallace:1971zz,Wallace:1973iu}  
that obtains the scattering propagator  as an expansion around the eikonal  propagator (Glauber approach) has the potential to
extend the range of validity of the approximation to lower
energies. Here we examine the properties of this expansion, and
calculate the first-, second-, and third-order corrections for the
scattering of a spinless particle off of a ${}^{40}$Ca
nucleus, and for nuclear breakup reactions involving ${}^{11}$Be. We find
that, including these
corrections extends the lower bound of the range of validity of the 
down to energies of 
40 MeV. At that energy   the corrections provide as much as a 15\%
correction to certain processes.
\end{abstract}
\maketitle
\section{Introduction}

Ongoing and planned experiments using rare isotopes promise to further
our understanding of nuclei and their role in
astrophysics~\cite{msu}. Nuclear reaction  theory is needed both to 
interpret the data and to determine the necessary
experiments~\cite{Hansen:2003sn,gg08,Bertulani:2009mf}. Use of the
eikonal approximation (also known as Glauber theory~\cite{roy}) has
long been known as appealing procedure to simplify the calculations,
for medium and low energies see {\it
  e.g.}~\cite{Bertsch:1990zz,Ogawa:1992tf,AlKhalili:1996zz,Hencken:1996af,Aumann:2000zz}. This
technique has often been used to analyze experiments, see {\it e.g}
\cite{Parfenova:2000be,Licot:1997zz,Sauvan:2000hq} performed at
energies less than 100 MeV per nucleon. A computer program using the
eikonal approximation, described as being appropriate for knockout
reactions for energies between 30 and 2000 MeV per nucleon, has been
published~\cite{Bertulani:2006fe}.  However, as stated in the orignal article~\cite{roy}  the Glauber theory rests
on the approximation that the product of the wave number $k$  and the
range of the relevant potential $a$  satisfy \bea k\,a\gg1\eea and
that the magnitude of the scattering potential $V$  be very small
compared to the scattering energy, $E$, so that  \bea V/E\ll1.\eea For
a nucleon of energy 100 MeV and nucleus of radius $\approx3$  fm,
$k\,a\approx 6$, and $V/E\approx 1/2.$ It is far from  obvious that the
conditions for the accuracy of the Glauber approximation are
satisfied.  Moreover, it is not clear if the relevant distance appearing in the term $ka$ should be the nuclear radius or the nuclear diffuseness. If the latter, the beam energy must be higher for the eikonal approximation to be valid.
 It is therefore of interest to assess the accuracy of
Glauber theory and the lower limits on energy for which it may be
applied~\cite{Esbensen:2001mr}. In the following we treat the terms eikonal approximation and Glauber theory as synonymous.

The conclusions of Ref.~\cite{Esbensen:2001mr} have been
summarized~\cite{Hansen:2003sn} as showing that the eikonal
approximation is accurate to within a few percent for energies as low
as  20 MeV/nucleon.  This conclusion is based on a comparison between
the results of using the eikonal approximation  and a time-dependent
Schr\"{o}dinger equation.
The incoming projectile is treated as a bound state of a nucleon and a
core. The time-dependent Schr\"{o}dinger equation that includes the
dynamics of the interaction of the nucleon with the core as well as
the nucleon-target interaction was solved. We do not believe the
conclusion that  the eikonal approximation is valid at 20
MeV~\cite{Hansen:2003sn}  to be  a valid summary of the work of
Ref.~\cite{Esbensen:2001mr}. This is because the time-dependent
equation (their Eqs.(5,6)) treats the motion of the core of the
projectile as following the linear trajectory $\bfR=\bfb +{\bf v
}t$. In other words,  the eikonal approximation is used in the
time-dependent Schroedinger equation.  Thus the work contains  no
actual test of the eikonal approximation. However,
Ref.~\cite{Esbensen:2001mr}  does have the very useful result that the
interaction between the nucleon and the core that occurs during the
nuclear reaction can be neglected for energies as low as 20
MeV/nucleon.  Thus the so-called sudden approximation is justified, at
least for one particular state. However, the use of the eikonal
approximation has not been justified and the range of its validity has
not been fully determined. Thus the present paper is devoted to
studying the corrections to the eikonal approximation.

In this paper we assess the validity of the eikonal approximation by
computing the corrections~Sect.~\ref{sec:et}  to this approximation
for potential scattering~Sect.~\ref{sec:pot}, and for reactions
involving halo nuclei Sect.~\ref{sec:halo}. The principal tool is the
expansion developed by Wallace~\cite{Wallace:1971zz,Wallace:1973iu} in
which the complete Green's function is expanded about the Glauber
approximation to the complete Green's function. Our results and
directions for further research are summarized in a final
Sect.~\ref{sec:summ}.

\section{Corrections to the Eikonal Theory}
\label{sec:et}
We first apply the corrections to the eikonal approximation
described by Wallace
\cite{Wallace:1971zz,Wallace:1973iu} for scattering of a spin-zero
particle off a generic potential. This exercise is useful because we
can calculate the scattering amplitude exactly using a partial wave
expansion and compare it with successive corrections in the eikonal
expansion. We will give a quick review of the corrections here using
the same notation as~\cite{Wallace:1971zz,Wallace:1973iu} before
showing the results of our calculations.

The  $T$ matrix for scattering at a center of mass energy $E=K^2/2M$  is given by 
\bea T(E)=V+VG_0(E)T=V+VG(E)T(E),\eea
where $G^{-1}(E)=E-P^2/2M-V+i\epsilon$ is the particle propagator and $V$ is the interaction potential.

The Wallace eikonal expansion consists of  expanding the momentum
operator $\mathbf{P}$ about a particular vector $\mathbf{k}$ and dropping all terms
quadratic in ${\bf P}-\mathbf{k}$. The choice $\bfk=K \widehat{\bfk}$
with $\widehat{\bfk}$ as the average of the projectile initial and
final direction ($\mathbf{k} = (\mathbf{k}_i + \mathbf{k}_f)/2$) gives
the Glauber approximation and the propagator:
\bea g^{-1}=\mathbf{v}\cdot(\mathbf{k}-\mathbf{P})-V+i\epsilon.\eea
The difference between the full propagator $G$ and the reduced eikonal
propagator $g$ is given by
\bea &&g^{-1}-G^{-1}=N\\
&& N=(1-\cos(\theta/2))(g^{-1}+V)+[(\mathbf{P}-\mathbf{k}_f)\cdot
(\mathbf{P}-\mathbf{k}_i)]/2M,\eea
where $\theta$ is the scattering angle.

It is then possible to solve for the $T$ matrix as a perturbation series:
\bea T=(V+VgV)+VgNgV+VgNgNgV+VgNgNgNgV+\dots.\eea

The Glauber approximation consists of keeping only the terms in
parentheses, and Wallace showed how to systematically calculate higher
order correction terms. The result is an expansion in powers of the
interaction energy over the kinetic energy with corrections due to the
spatial non-uniformity of the potential. He explicitly calculates the
first three correction terms, and first with the conjecture of some
advantageous cancellations~\cite{Wallace:1971zz,Wallace:1973iu}, and later~\cite{Wallace:1973ni}
in an explicit calculation
 obtained the following expressions:
\begin{align}
\label{Eq:Tmats}
T^{(0)}(\mathbf{b}) & = e^{i\chi_0(|\mathbf{b}|)}-1\\
T^{(1)}(\mathbf{b}) & = e^{i(\chi_0(|\mathbf{b}|)+\tau_1(|\mathbf{b}|))}-1\\
T^{(2)}(\mathbf{b}) & = e^{i(\chi_0(|\mathbf{b}|)+\tau_1(|\mathbf{b}|)+\tau_2(|\mathbf{b}|))}e^{-\omega_2(|\mathbf{b}|)}-1\\
T^{(3)}(\mathbf{b}) & = e^{i(\chi_0(|\mathbf{b}|)+\tau_1(|\mathbf{b}|)+\tau_2(|\mathbf{b}|) +\tau_3(|\mathbf{b}|) +\phi_3(|\mathbf{b}|))} e^{-\omega_2(|\mathbf{b}|) -\omega_3(|\mathbf{b}|)}-1.
\end{align}
Here $\mathbf{b}$ is the impact parameter, $T^{(0)}$ is the Glauber
approximation, and the phases are defined below, with
$\mathbf{z}\perp\mathbf{b}$, $\mathbf{r}=\mathbf{b} +
\mathbf{z}$, $U(r) = V(r)/V(0)$, $\widehat{\beta}_n \equiv
b^n \partial^n/\partial b^n$, and $\epsilon=V(0)/2E$:
\begin{align}
\chi_0(b) & = -2K\epsilon \int_0^\infty \! \mathrm{d}z \, U(r)\\
\tau_1(b) & = -K\epsilon^2(1+\widehat{\beta}_1) \int_0^\infty \! \mathrm{d}z \, U^2(r)\\
\tau_2(b) & = -K\epsilon^3(1+\frac53\widehat{\beta}_1+\frac13\widehat{\beta}_2)
\int_0^\infty \! \mathrm{d}z \, U^3(r)-\frac{b[\chi_0'(b)]^3}{24K^2}\\
\omega_2(b) & = b\chi_0'(b)\frac{\nabla^2\chi_0(b)}{8K^2}\\
\tau_3(b) & = -K\epsilon^4(\frac54+\frac{11}{4}\widehat{\beta}_1+\widehat{\beta}_2+\frac{1}{12}\widehat{\beta}_3)\int_0^\infty \! \mathrm{d}z \, U^4(r)-\frac{b\tau_1'(b)[\chi_0'(b)]^2}{8K^2}\\
\phi_3(b) & = -K\epsilon^2(1+\frac53\widehat{\beta}_1+\frac13\widehat{\beta}_2) \int_0^\infty \! \mathrm{d}z \, \left[\frac{\partial U(r) / \partial r}{2K}\right]^2\\
\omega_3(b) & =
\frac{b\chi_0'(b)\nabla^2\tau_1(b)+b\tau_1'(b)\nabla^2\chi_0(b)}{8K^2}
\label{Eq:phases}
\end{align}
We see that the corrections related to $\widehat{\beta}_n$ involve the derivatives of the nuclear potential which  are large in the region of the  nuclear surface. This  indicates that the product of the wave number and the nuclear diffuseness parameter, needs to be large  compared to unity for the   eikonal approximation to be valid.  This condition is  more stringent than the one involving  the product of the wave number and the nuclear radius.

The scattering amplitude is then simply:
\bea f(\mathbf{q}) = -iK/2 \int \! \mathrm{d^2}b \, e^{i {\bf q \cdot b}
} T^{(n)}(\mathbf{b}).\eea

\section{Eikonal Expansion {\it vs.} Exact Partial Wave Results}
\label{sec:pot}
Our focus is on reactions at FRIB energies. We therefore evaluate the
scattering amplitude for protons scattering off of ${}^{40}$Ca using
the potential described by Varner et. al.~\cite{Varner:1991zz} for
incident center-of-mass kinetic energy between 16 and 98 MeV. We
neglect the spin-orbit  and the Coulomb interaction because such terms
are neglected in Ref.~\cite{Hencken:1996af}.
 
 This potential is then given by 
 \bea
V(r, E)=-V_r(E)\,f_{\rm ws}(r,R_0,a_0)-i W_{\rm v}(E) \,f_{\rm ws}(r,R_{\rm w},a_{\rm w})-i\,W_{\rm s}(E){d\over dr}\,f_{\rm ws} (r,R_{\rm w},a_{\rm w}),\eea
with
\bea
f_{\rm ws}(r,R,a) & = & {1\over 1+\exp[(r-R)/a]}\\
V_r(E) & = & V_0 + V_{\mathrm{e}}(E-E_{\mathrm{c}}) \pm V_{\rm t}\varepsilon\\
W_{\mathrm{v}}(E) & = & W_{\rm v0}f_{\mathrm{ws}}(W_{\mathrm{ve0}},
(E-E_{\mathrm{c}}), W_{\mathrm{vew}})\\
W_{\mathrm{s}}(E) & = & (W_{\rm s0} \pm W_{\rm
  st}\varepsilon)f_{\mathrm{ws}}((E-E_{\mathrm{c}}), W_{\mathrm{se0}},
W_{\mathrm{sew}}),
\eea
where the $\pm$ indicates $+$ for proton projectiles and $-$ for
neutron projectiles. Parameters in the model can be found in
Table~\ref{tab:params} and in the text.
\begin{table}[h]
\centering
  \begin{tabular}{ l l l }
    \hline
    Parameter & Value & Uncertainty \\
    \hline
    $\varepsilon$ & $(N-Z)/A$ & -- \\
    $V_0$ & 52.9 MeV & $\pm~0.2$ \\   
    $V_{\rm t}$ & 13.1 MeV & $\pm~0.8$ \\
    $V_{\rm e}$ & -0.299 & $\pm~0.004$ \\
    $r_0$ & 1.250 fm & $\pm~0.002$ \\
    $r_0^{(0)}$ & -0.225 fm & $\pm~0.009$ \\
    $R_0$ & $r_0A^{1/3} + r_0^{(0)}$ & -- \\
    $a_0$ & 0.690 fm & $\pm~0.006$ \\
    $r_{\rm c}$ & 1.24 fm & -- \\
    $r_{\rm c}^{(0)}$ & 0.12 fm & -- \\
    $R_{\rm c}$ & $r_{\rm c}A^{1/3}+r_{\rm c}^{(0)}$ fm & -- \\
    $E_{\mathrm{c}}$ & $6Ze^2 / 5R_{\mathrm{c}}$ MeV& -- \\
    $W_{\rm v0}$ & 7.8 MeV & $\pm~0.3$ \\
    $W_{\rm ve0}$ & 35 MeV & $\pm~1$ \\
    $W_{\rm vew}$ & 16 MeV & $\pm~1$ \\
    $W_{\rm s0}$ & 10.0 MeV & $\pm~0.2$ \\
    $W_{\rm st}$ & 18 MeV & $\pm~1$ \\
    $W_{\rm se0}$ & 36 MeV & $\pm~2$ \\
    $W_{\rm sew}$ & 37 MeV & $\pm~2$ \\
    $r_{\rm w}$ & 1.33 fm & $\pm~0.01$ \\
    $r_{\rm w}^{(0)}$ & -0.42 fm & $\pm~0.03$ \\
    $R_{\rm w}$ & $r_{\rm w}A^{1/3} + r_{\rm w}^{(0)}$ fm& -- \\
    $a_{\rm w}$ & 0.69 fm & $\pm~0.01$ \\
    \hline
  \end{tabular}
\caption{Parameters of interest from ref~\cite{Varner:1991zz}}
\label{tab:params}
\end{table}

The approximate eikonal solution to potential scattering can be
compared with an exact solution obtained by using a partial wave
technique. Phase shifts for arbitrary values of $\ell$ are obtained
by numerically solving the radial Schr\"{o}dinger equation,
$$\frac{d^2u_\ell}{dr^2}+\left(k^2-2mV-\frac{\ell(\ell+1)}{r^2}\right)u_\ell=0,$$
matching $u_\ell(R) = Re^{i\delta_\ell}[\cos(\delta_\ell)j_\ell(kR)-\sin(\delta_\ell)n_\ell(kR)]$ for $R$ large enough such that $V(R) \approx 0$, and solving for $\delta_\ell$. Here, $j_\ell$ and $n_\ell$ are spherical Bessel functions of the first and second kind, respectively. The scattering amplitude is then:
$$f(\theta) = \sum_{\ell=0}^\infty (2\ell+1)e^{i\delta_\ell}\sin\delta_\ell P_\ell(\cos\theta).
$$
This sum is taken to convergence and compared to scattering amplitudes
derived from the eikonal approximation. This comparison provides an
indication of the validity of the approximation when applied to more
complex systems for which an exact solution is not feasible. 
\begin{figure}[h]
\centering
\includegraphics[width=\textwidth]{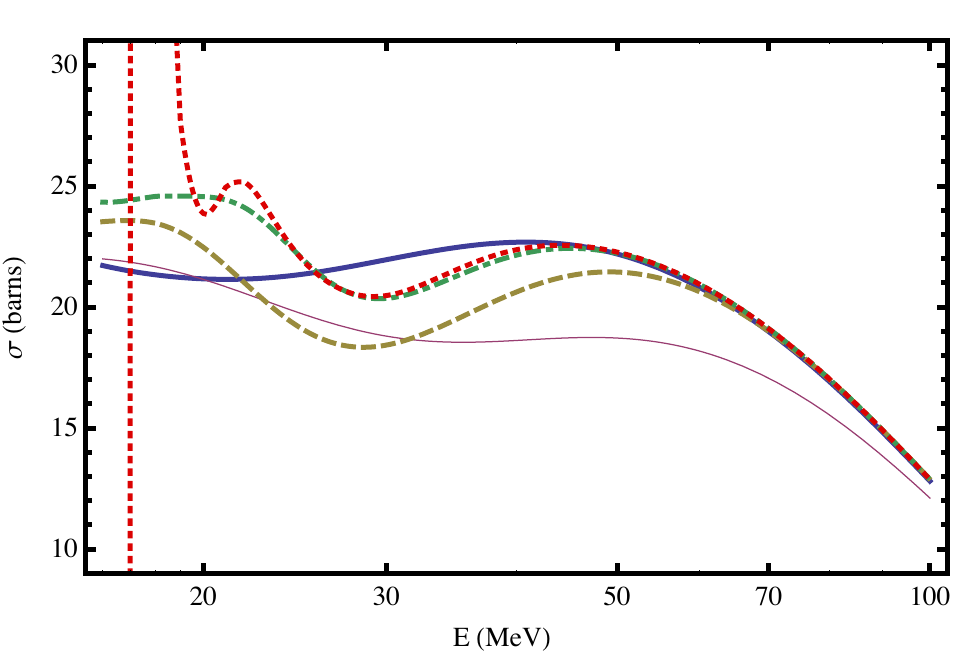}
\caption{\label{fig:totalcs}(Color online) Total nuclear
  cross-section $\sigma$ for a proton incident on ${}^{40}$Ca as a
  function of beam energy. The exact partial wave result is the thick
  blue line, the zeroth-order eikonal approximation is
  the thin magenta line, the first-order eikonal approximation is the
  beige dashed line, the second-order eikonal approximation is the
  green dot-dashed line, and the third-order eikonal approximation is
  the red dotted line.}
\end{figure}
\subsection{Results of Calculations}
\label{subsec:potres}
\begin{figure}[h]
\centering
\includegraphics[width=\textwidth]{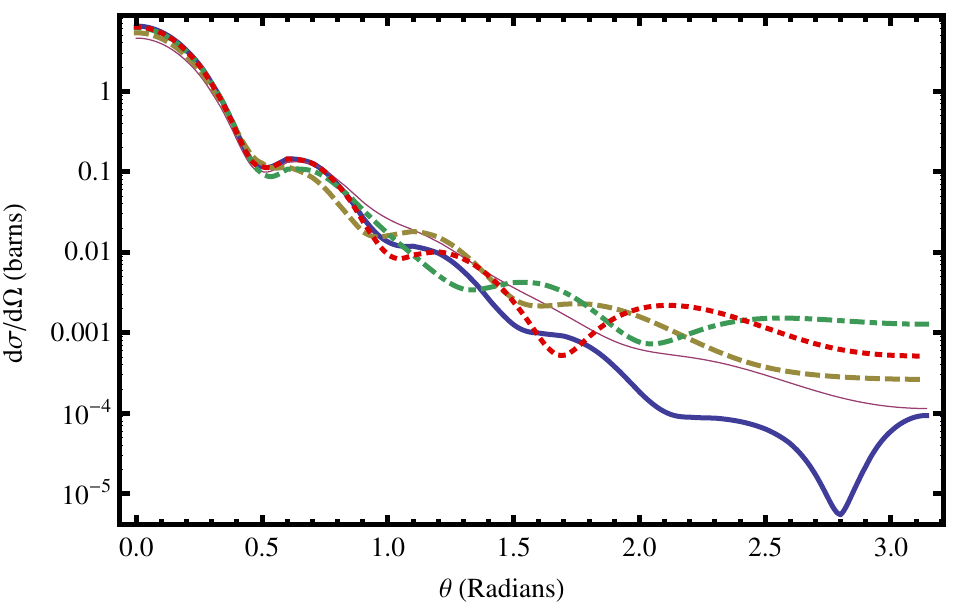}
\caption{\label{fig:logdiffcs} (Color online) Differential elastic
  cross-section $d\sigma/d\Omega$ for p +
  ${}^{40}$Ca at a beam energy of 40 MeV in log scale. The
  angle $\theta$ is the scattering angle from the forward direction. The
  designations for the lines are the same as in Fig.~\ref{fig:totalcs}.}
\end{figure}

\begin{figure}[h]
\centering
\includegraphics[width=\textwidth]{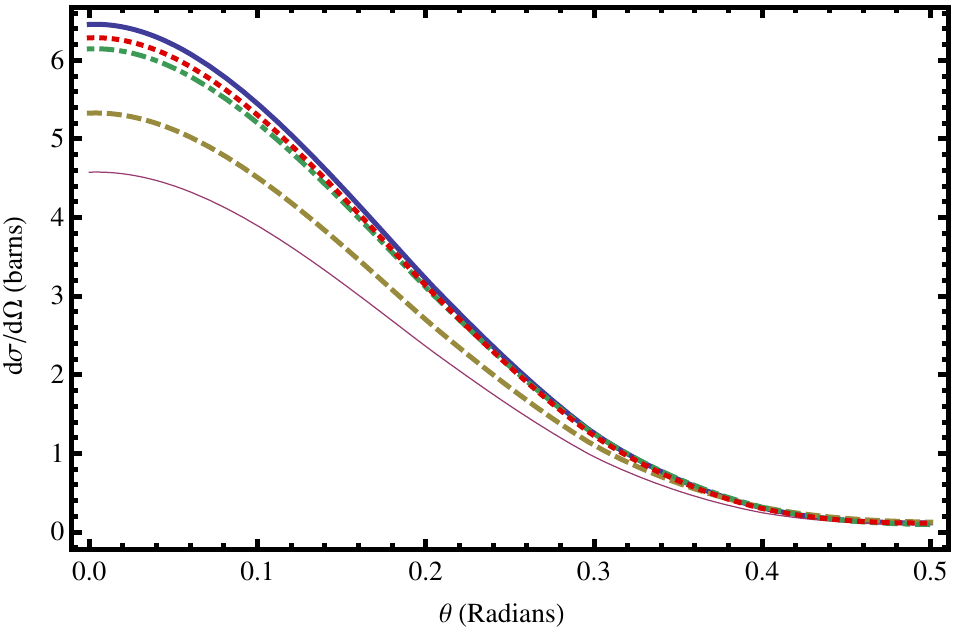}
\caption{\label{fig:lindiffcs} (Color online) Same as
  Fig.~\ref{fig:logdiffcs}, but zoomed in on the forward scattering
  region, and with a linear scale.}
\end{figure}

\begin{figure}[h]
\centering
\begin{subfigure}{0.45\textwidth}
\includegraphics[width=\textwidth]{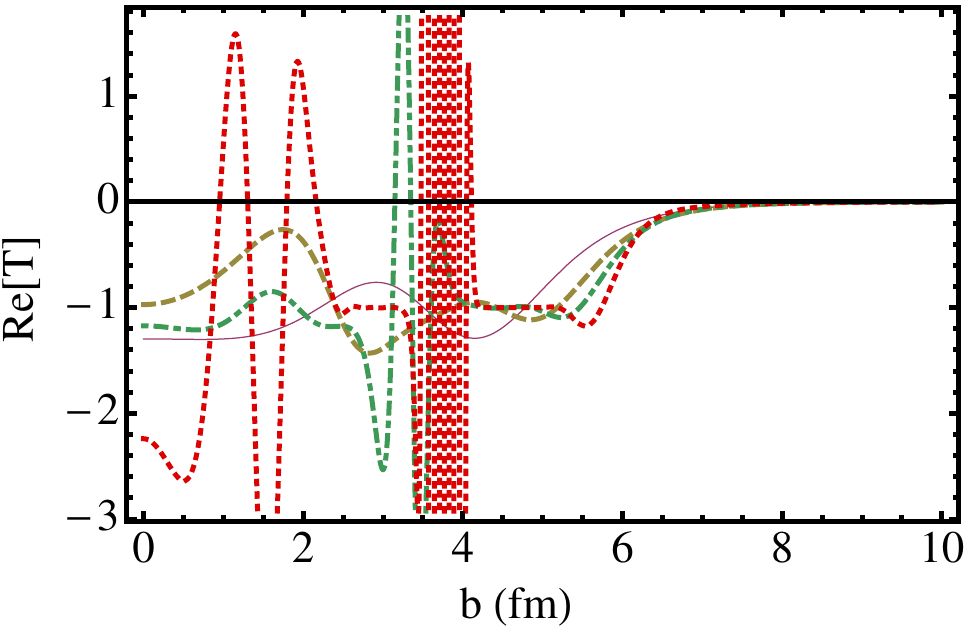}
\caption{Beam Energy at 20 MeV}
\label{fig:tfnsa}
\end{subfigure}
\begin{subfigure}{0.45\textwidth}
\includegraphics[width=\textwidth]{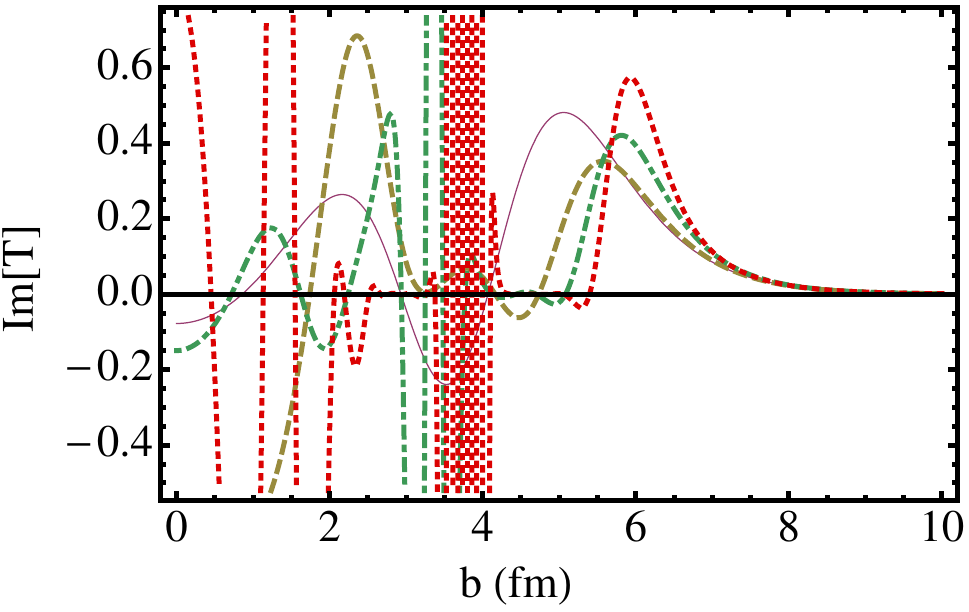}
\caption{Beam Energy at 20 MeV}
\end{subfigure}
\begin{subfigure}{0.45\textwidth}
\includegraphics[width=\textwidth]{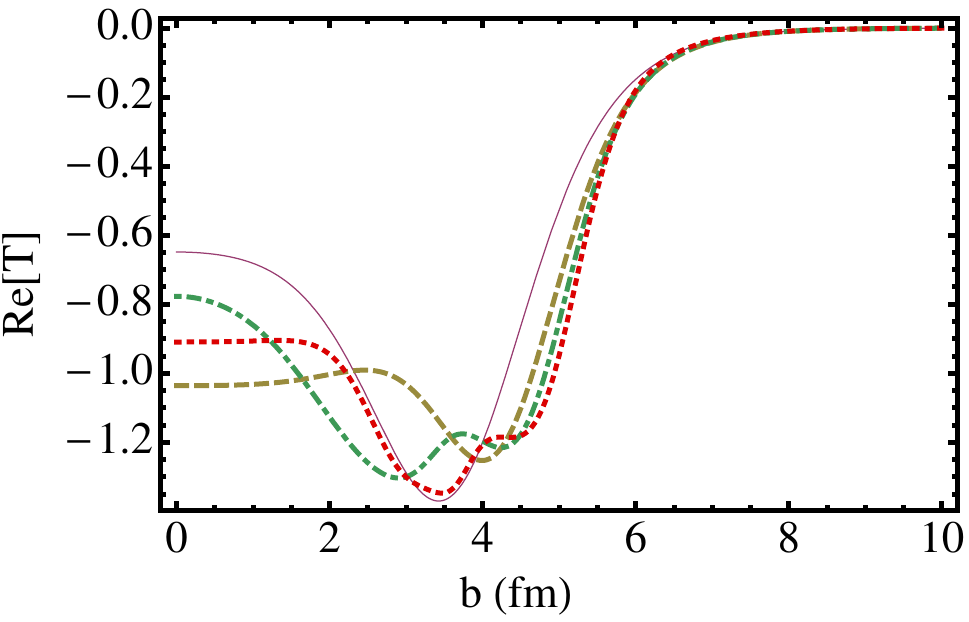}
\caption{Beam Energy at 40 MeV}
\end{subfigure}
\begin{subfigure}{0.45\textwidth}
\includegraphics[width=\textwidth]{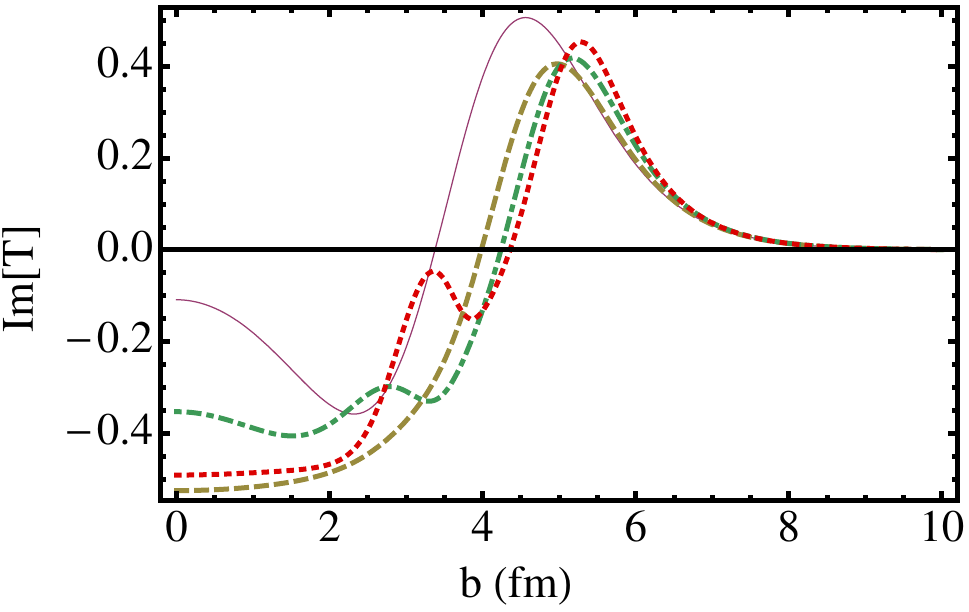}
\caption{Beam Energy at 40 MeV}
\end{subfigure}
\begin{subfigure}{0.45\textwidth}
\includegraphics[width=\textwidth]{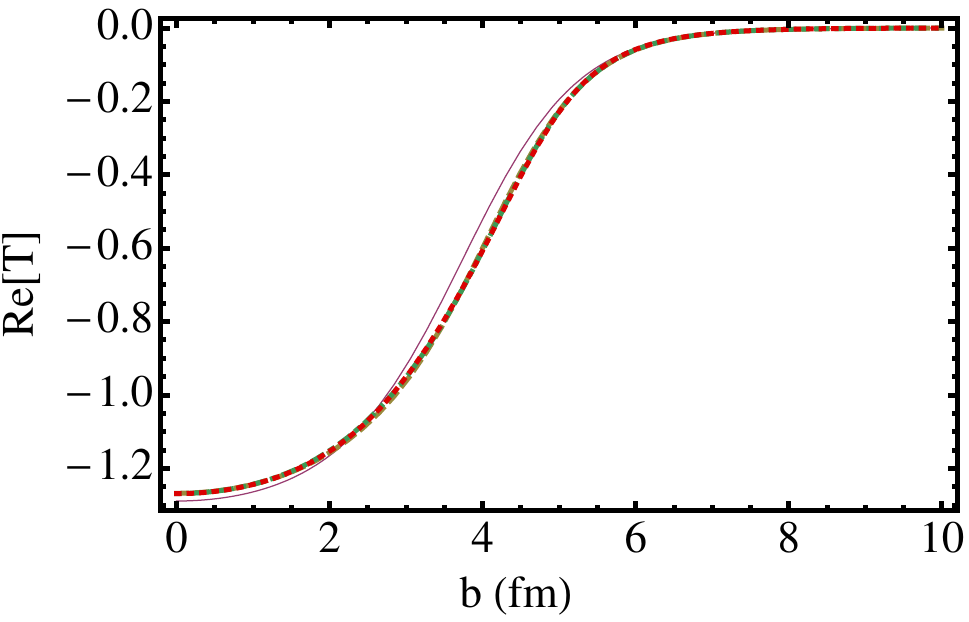}
\caption{Beam Energy at 98 MeV}
\end{subfigure}
\begin{subfigure}{0.45\textwidth}
\includegraphics[width=\textwidth]{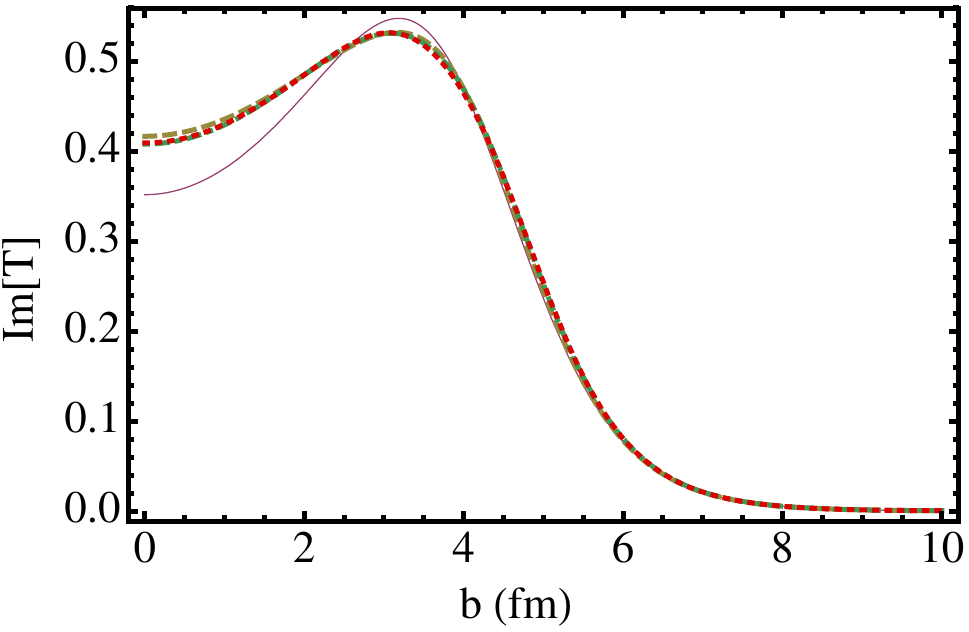}
\caption{Beam Energy at 98 MeV}
\end{subfigure}
\caption{(Color online) Real and imaginary parts of the transition
  matrix elements $T^{(n)}(b)$ for successive orders $n$ in the
  eikonal expansion. The designations for the lines are the same as in
  figs.~\ref{fig:totalcs}.}
\label{fig:tfns}
\end{figure}

Our main results for these calculations are presented in
figures~\ref{fig:totalcs}-\ref{fig:tfns}. Figure~\ref{fig:totalcs} shows the total
elastic nuclear cross-section of a spinless proton incident on a
${}^{40}$Ca nuclear potential in the exact calculation, and in
successive orders in the eikonal expansion. The zeroth order eikonal
approximation has an error of at least 5\% up to 100 MeV, while
including the correction terms reduces the error to $<$ 1\% above
about 45 MeV. The relative degree of agreement between the zeroth
order approximation and the exact calculation at energies below 20 MeV
is likely a coincidence.


Figures~\ref{fig:logdiffcs} and~\ref{fig:lindiffcs} show the
differential elastic cross-section for the same
reaction at a beam energy of 40 MeV. Even for
forward scattering, the zeroth-order eikonal approximation severly
underestimates the exact value, and successive corrections
monotonically improve the estimate. The
corrections also successively improve the range in the polar angle
$\theta$ over which the approximation is accurate.

\begin{figure}[h]
\centering
\begin{subfigure}{0.45\textwidth}
\includegraphics[width=\textwidth]{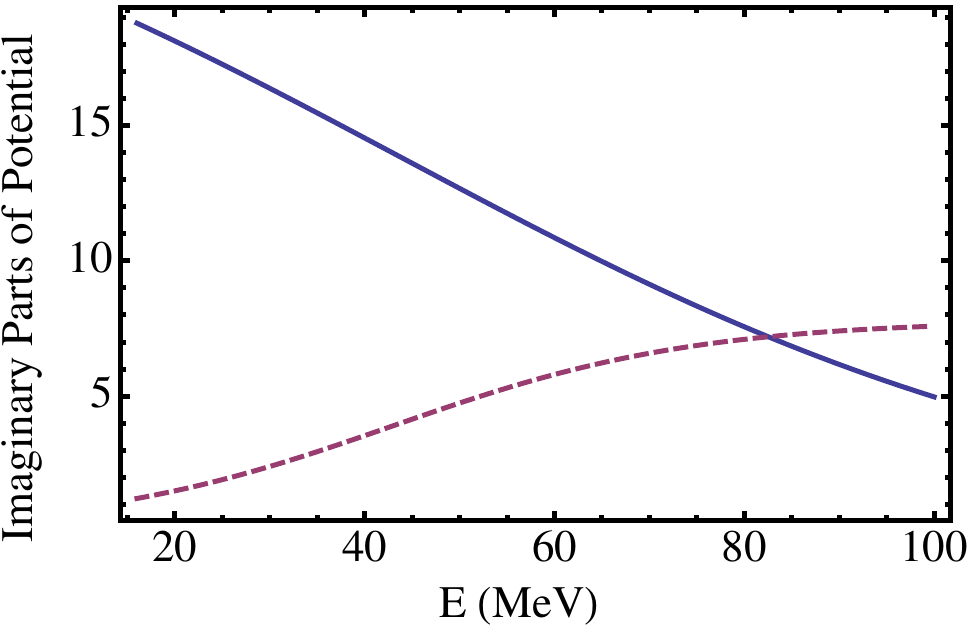}
\caption{Imaginary surface (solid blue) $W_{\rm s}(E)$ and volume
  (dashed magenta) $W_{\rm v}(E)$ terms of Varner potential as a function of beam
  energy.}
\label{fig:WvWs}
\end{subfigure}
\begin{subfigure}{0.45\textwidth}
\includegraphics[width=\textwidth]{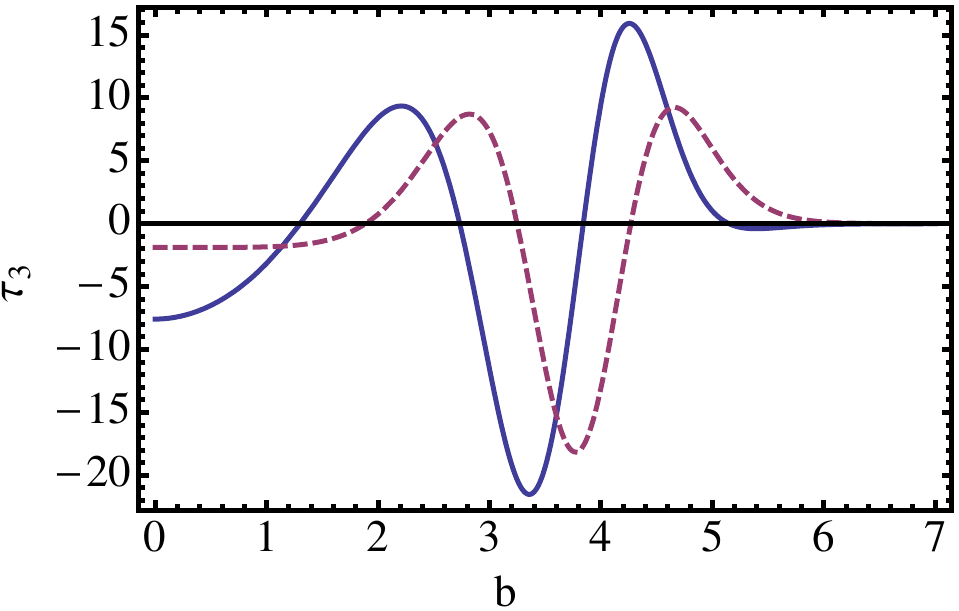}
\caption{Real (solid blue) and imaginary (dashed
  magenta) parts of $\tau_3(b)$ at a beam energy of 20 MeV.}
\label{fig:tau320}
\end{subfigure}
\caption{(Color online)}
\label{fig:plots}
\end{figure}
\begin{figure}
\begin{subfigure}{0.45\textwidth}
\includegraphics[width=\textwidth]{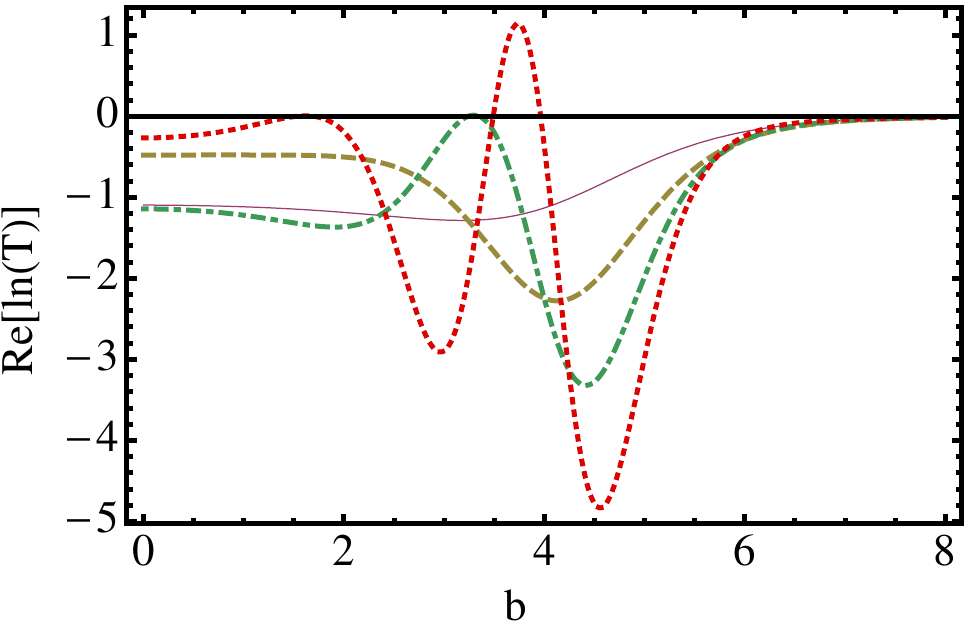}
\caption{}
\label{fig:relnT25}
\end{subfigure}
\begin{subfigure}{0.45\textwidth}
\includegraphics[width=\textwidth]{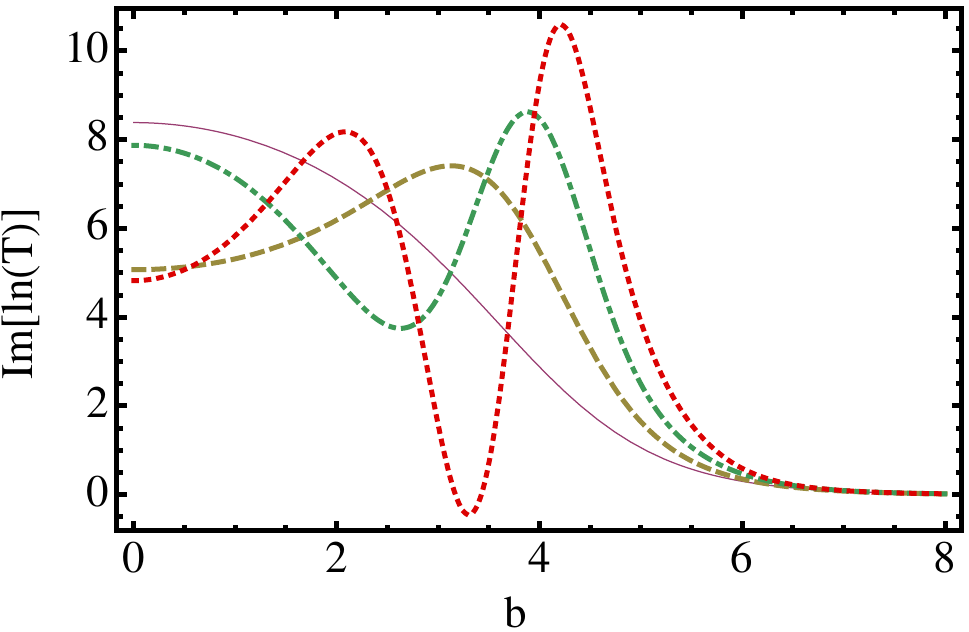}
\caption{}
\label{fig:imlnT25}
\end{subfigure}
\caption{(Color online) Real and imaginary parts of $\ln(T^{(n)}(b))$ for
  $n=0$ (thin magenta), $n=1$ (dashed beige), $n=2$ (dot-dashed
  green), and $n=3$ (dotted red) calculated at a beam energy of 25
  MeV.}
\label{fig:lnT25}
\end{figure}
Figure~\ref{fig:tfns} gives the real and imaginary parts of the $T$-matrix elements $T^{(n)}(b)$ for successive orders $n$ in the expansion as a
function of the impact parameter $b$ at a variety of beam
energies. The rapid oscillations and drastic changes in $T$ with each
correction at 20 MeV imply that the expansion is not appropriate
there. This is because the interaction potential is energy dependent. In this
case, it is the imaginary part of the potential that is important. It
has both a surface and a volume term which have magnitudes that behave
oppositely as a function of beam energy, as shown in
fig.~\ref{fig:WvWs}. At low energies, the surface term dominates and
the derivative operators $\widehat{\beta}_n$
in $\tau_3(b)$ are large, negative, and imaginary, (see
Fig.~\ref{fig:tau320}) which generate
large oscillations in $T^{(3)}(b)$. (The frequencies
of such oscillations are given by the real part of $\tau_3(b)$.) This
behavior also occurs in $\tau_2(b)$ and $\tau_1(b)$, but at lower
energies. The point at which this breakdown occurs provides a lower
bound on the effectiveness of the expansion that can be computed
order-by-order. For example, in fig.~\ref{fig:relnT25}, which was
calculated at 25 MeV, the third-order correction has a real part of
about 1, which is already an amplitude of oscillations in
$T^{(3)}(b)$ of about 2.7 at $b \approx 3.75$. The second-order correction is about
to enter positive territory in fig.~\ref{fig:relnT25} at $b \approx
3.25$, and will start to generate similar rapid oscillations in
$T^{(2)}(3.25)$ at lower eneries. This can be seen in
fig.~\ref{fig:tfnsa}, which was calculated at 20 MeV. Thus,
empirically, the second-order correction is effective to about 25 MeV
for this potential. Using the same method, we find the third-order
correction to be effective to about 30 MeV. 

Since the convergence of the expansion improves at higher energies,
calculating only the first-order correction should be sufficient at
some sufficiently high beam energy. From fig.~\ref{fig:totalcs} this
appears to happen for this potential at a beam energy of about 60
MeV. Above this value, the fractional error in the second- and
third-order corrections is only marginally lower than the fractional
error in the first-order correction.

With these calculations, it is apparent that for at least some
interactions, these corrections to the eikonal approximation are meaningful
over a range of energies. It is therefore worthwhile to apply the
corrections to a more interesting interaction to further evaluate
their effectiveness.


\section{Breakup Reactions of Halo Nuclei ${}^{11}$Be}
\label{sec:halo}
We now apply these calculations to the study of scattering of
${}^{11}$Be off of various targets, using the reaction theory of
Hencken, Bertsch \& Esbensen~\cite{Hencken:1996af}. They computed the
diffractive, neutron stripping, core stripping, and total absorption
cross sections for ${}^{11}$Be scattered off targets with mass number
ranging from 9-208 using the Glauber eikonal approximation at an
energy of 40 MeV/nucleon. They used the Varner
potential~\cite{Varner:1991zz} as the model for nucleon-nucleon
scattering. Given that we see a significant improvement in the
performance of the eikonal approximation at that energy when the
Wallace corrections are included for the Varner potential, it is
fruitful to investigate whether or not the cross-sections evaluated by
Hencken and Bertsch also experience similar improvement.

\begin{figure}[h]
\includegraphics[width=7.8cm,height=5.5cm]{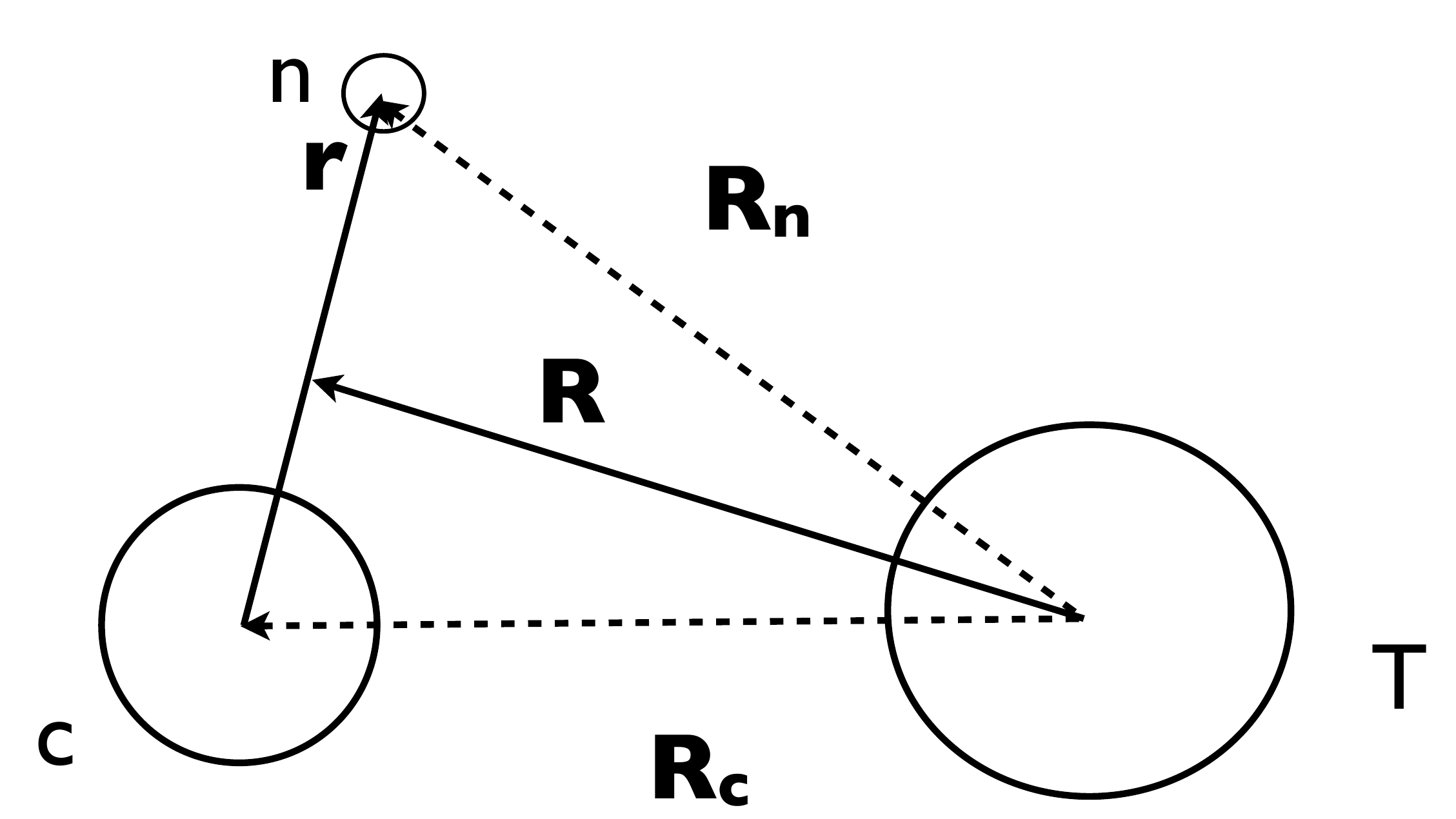}
\caption{Coordinates used in this calculation. ${\bf R}$ is the
  coordinate of the center of mass of the halo nucleus, and ${\bf b}_c$
  and $\bfb_n$ denote the components of $\mathbf{R}_c$ and $\mathbf{R}_n$
  that are transverse to the beam direction.}\label{fig:coords}\end{figure}

The relevant formulae of Ref.~\cite{Hencken:1996af} are displayed next.
The reaction considered is $H +T\rightarrow c+X$, where the
projectile halo nucleus $H$  is treated in a single particle model as
$c+n$ with $c$ corresponding to a specific final state of the
core. The halo nuclear ground state is described by a wave function
$\phi_{LM}(\bfr)$ which depends on the relative coordinate $\vec{r}$
between the nucleon and the core, see Fig.~\ref{fig:coords}. The function
is generally specified by $\phi_{LM}(\bfr)=R_L(r) Y_{LM}(\widehat{\bfr})$ where
 $Y_{LM}(\widehat{\bfr})$ are   spherical harmonics. Here we take $R_L(r)$
to be  the solution to the  radial Schr\"{o}dinger equation in an $L=0$ state
with
the appropriate binding energy of 0.503 MeV.

The scattering wave function of the halo nucleus has the form,
\begin{equation}
\Psi(\bfr,\bfR) = S_n(\bfb_n) S_c(\bfb_c) \phi_{LM}(\bfr) 
\ ,
\label{Eq:final}
\end{equation}
in its rest frame, where (Fig.~\ref{fig:coords}) $\bfR$ is the coordinate of the
center of mass of the halo 
nucleus, and $\bfb_c$ and $\bfb_n$ are the impact parameters of
the core and the nucleon with respect to the target nucleus, i.~e.
$\bfb_n=\bfR_\perp + \bfr_\perp A_c/(A_c+1)$ and
$\bfb_c=\bfR_\perp - \bfr_\perp /(A_c+1)$, where $A_c$ is the
mass number of the core and the designation $\perp$ refers to components transverse to $\bfR_n$ and $\bfR_c$.  The two profile functions, $S_n(\bfb_n)$
for the nucleon and $S_c(\bfb_c)$ for the core, are generated by
interactions with the target nucleus.  In the eikonal approximation,
they are defined by the longitudinal integrals over the corresponding
potentials:
\begin{equation}
S(\bfb) = \exp \left[\frac{-i}{\hbar v}
\int dz V(\bfb + z  \widehat{z})  \right] \ ,
\label{Eq:profile}
\end{equation}
where $v$ is the beam velocity and  potential $V$ is the optical
potential.
The relation between $S(\bfb)$ and the quantities denoted as $T(\bfb)$
of Sect.~\ref{sec:et}) is given by
\bea S(\bfb)=T(\bfb)+1.\eea
We compute the order $n$ corrections by replacing $S(\bfb)$ from
equation~\ref{Eq:profile} with $T^{(n)}(\bfb) + 1$ from
equations~\ref{Eq:phases}.

The scattering wave function is the
difference between eq.~(\ref{Eq:final}) and the wave function of the
undisturbed beam,
\begin{equation}
\Psi_{scat} = (S_n S_c -1 ) \phi_{LM} \ .
\end{equation}
with the shorthand notation $S_n=S_n(\mathbf{b_n})$ and $S_c=S_c(\mathbf{b_c})$.

Scattering cross sections are calculated by taking overlaps
of $\Psi_{scat}$ with different final states.  
For diffractive breakup the final state depends on the relative
momentum $\vec{k}$ of nucleon and core in their center-of-mass frame
as well as on the transverse momentum $\vec{K}_\perp$ of the center
of mass. Writing the continuum nucleon-core wave function as
$\phi_{\bfk}(\vec{r})$ (normalized asymptotically to a plane wave: $
\phi_{\bf k}\sim\exp(i\bfk\cdot\bfr) $) the diffractive breakup cross
section is given by
\begin{eqnarray}
&&\frac{d\sigma_{\text{diff.}}}{\left(d^2\ {K}_{\perp} d^3{k}\right)}= \frac{1}{(2\pi)^5} \frac{1}{2 L+1} \sum_{M}
\left|
\int d^3{r} d^2{R}_{\perp} 
e^{-i{\bf K}_\perp\cdot{\bfR}_\perp} 
\phi_{{\bf k}}^*(\bfr) S_c S_n \phi_{LM}(\bfr) \right|^2 \ .
\end{eqnarray}
To obtain the relative momentum distribution   in
$\vec{k}$, integrate over ${\bf K}_{\perp}$ to get
\begin{eqnarray}
\frac{d\sigma_{\text{diff.}}}{d^3 {k}} 
&=& \frac{1}{(2\pi)^3} \frac{1}{2 L+1} \sum_{M} 
\int d^2{R}_\perp \left|
\int d^3{r} \phi_{{\bf k}}^*(\bfr) S_c S_n \phi_{L,M}
(\bfr) \right|^2\ .
\label{Eq:DDiff}
\end{eqnarray}

A convenient expression for the total diffractive cross section can be
derived using completeness if $\phi_{LM}$ is the only bound state of
the system. The result is
\begin{eqnarray}
\sigma_{\text{diff.}} &=& \frac{1}{2 L+1} \sum_{M}
\int d^2{R}_\perp  
\Bigg[\int d^3\vec{r} \phi_{L,M}(\bfr)^* 
\left|S_c S_n\right|^2 \phi_{L,M}(\bfr) 
- \sum_{M}
\left|\int d^3r \phi_{0,M_0'}(\bfr)^* S_c S_n 
\phi_{L,M}(\bfr) \right|^2 \Bigg] \ . \nonumber\\
&&\label{Eq:diff}
\end{eqnarray}

Other contributions to the total cross section come from absorption,
present when the eikonal $S$-factors have moduli less than 1.  There are
three of these so-called stripping processes.  The nucleon-absorption
cross section, differential in the momentum of the core, is given by
\begin{eqnarray}
\frac{d\sigma_{\text{n-str.}}}{d^3{k}_c} 
&=& \frac{1}{(2\pi)^3} \frac{1}{2 L+1} \sum_{M} \int d^2{b}_n 
\left[1- \left|S_n(\bfb_n)\right|^2 \right] \times 
\left| \int d^3\bfr e^{-i {\bf k}_c\cdot\bfr} 
S_c(\bfb_c) \phi_{L,M}(\bfr) \right|^2 \ .
\end{eqnarray}
The corresponding total cross section for stripping of the nucleon is 
\begin{eqnarray}
\sigma_{\text{n-str.}} &=& \frac{1}{2 L +1} \sum_{M }
\int d^2 {b}_n
\left[1- \left|S_n(\bfb_n) \right|^2 \right]  \int d^3\bfr \phi_{L,M}(\bfr)^* 
\left|S_c(\bfb_c)\right|^2 \phi_{L,M}(\bfr) \ .
\label{Eq:strip}
\end{eqnarray}
The stripping of the core is expressed in a similar way, interchanging
subscripts $n$ and $c$.

The expression for absorption of both nucleon and core is
given by
\begin{eqnarray}
\sigma_{\text{abs.}} &=& \frac{1}{2 L +1} \sum_{M }
\int d^2{b}_c 
\left[ 1 - \left| S_c(\bfb_c) \right |^2\right] \times \int d^3\vec{r} \phi^*_{L,M}(\bfr) 
\left[ 1 - \left| S_n(\bfb_n) \right |^2\right] 
\phi_{L,M}(\bfr) \ .
\label{Eq:absorb}
\end{eqnarray}

\subsection{The potential for the Nucleon-Target and Core-Target 
Interaction}
\label{sec:potential}

Evaluation of the profile functions requires a potential model for the
interaction between the target nucleus and the constituents of the
halo nucleus.  At low energies, extending up to about 100 MeV/$n$, one
can find optical potentials that are fit to nucleon-nucleus
scattering.  We use the optical potential, $V_{\rm op}$ of
ref.~\cite{Varner:1991zz}, which was fit to scattering data in the
range of 10 to 60 MeV.  The potential has the usual Woods-Saxon form,
with volume and surface imaginary terms, but we neglect the spin-orbit
and Coulomb interactions as does~\cite{Hencken:1996af}. This potential
represents the target-nucleon interaction. The core-target interaction
potential is obtained by folding $V_{\rm op}$ with the core density
distribution,
\begin{eqnarray}
V_c(r) = \int d^3 {x} \rho_c(x) 
 V_{\text{op}}(\left|\bfr - \bf{x}\right|) \ .
\end{eqnarray}
For the core density we use a harmonic oscillator density with
parameters taken from the charge distribution of the core nucleus
\cite{devries87} ($a$=2.5~fm and $\alpha$=0.61).

\subsection{Results of Eikonal Expansion Calculations}
\label{sec:res}
\begin{figure}[h]
\centering
\includegraphics[width=0.6\textwidth]{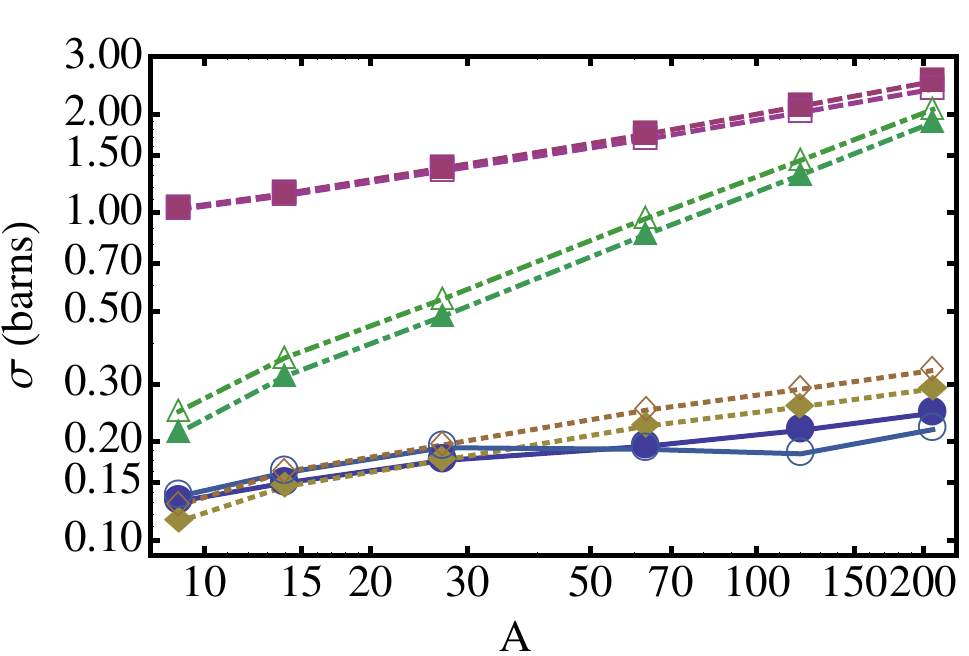}
\caption{(Color online) Comparison of first-order corrections with standard
  (zeroth-order) eikonal approximation with a beam energy of 40
  MeV/nucleon. The zeroth-order terms are shown with solid markers, and the
  first-order terms are with outlined markers. The solid (blue) line
  with circles is diffractive scattering, the dashed (magenta) line with
  squares is core stripping, the dotted (beige) line with diamonds is
  neutron stripping, and the dash-dotted (green) line with triangles is
  total absorption of the core and neutron.}
\label{fig:BertschPlot}
\end{figure}
\begin{figure}[h]
\includegraphics[width=\textwidth]{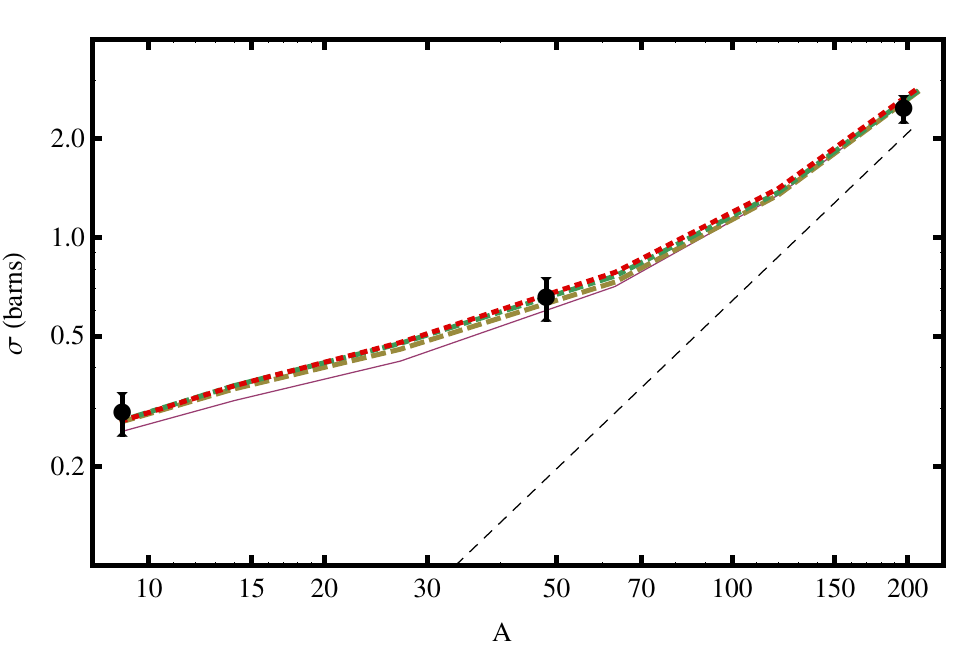}
\caption{(Color online) Comparison of corrections to scattering data
  at 41 MeV/nucleon from~\cite{Anne:1994}. The designations for the
  lines are the same as in fig.~\ref{fig:tfns}, with the Coulomb
  breakup cross-section in dashed black. Because we did not calculate
  corrections to the Coulomb cross-section, the corrections appear
  smaller on this plot at high energies where the Coulomb term is larger.}
\label{fig:compexp}
\end{figure}
\begin{figure}[h]
\centering
\begin{subfigure}{0.45\textwidth}
\includegraphics[width=\textwidth]{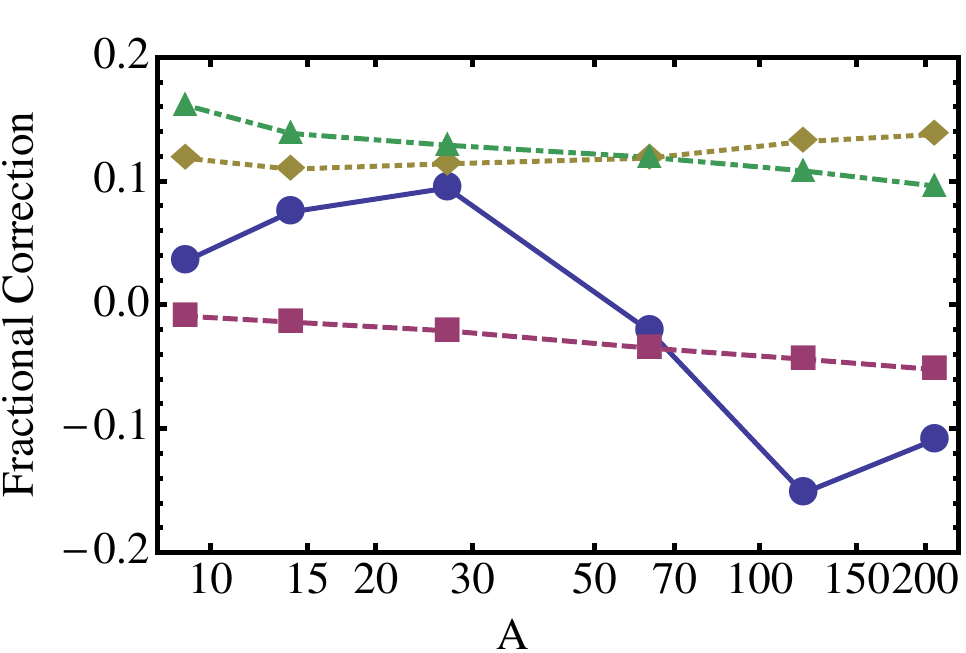}
\caption{Effect of first-order
  corrections at beam energy of 40 MeV.}
\end{subfigure}
\begin{subfigure}{0.45\textwidth}
\includegraphics[width=\textwidth]{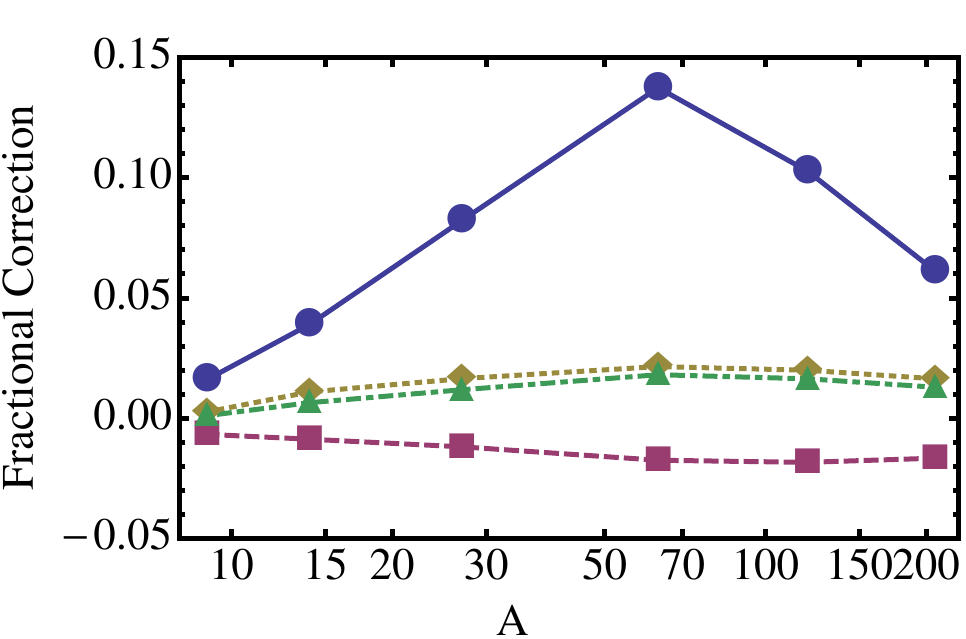}
\caption{Effect of second-order corrections at beam energy of 40 MeV.}
\end{subfigure}
 \begin{subfigure}{0.45\textwidth}
\includegraphics[width=\textwidth]{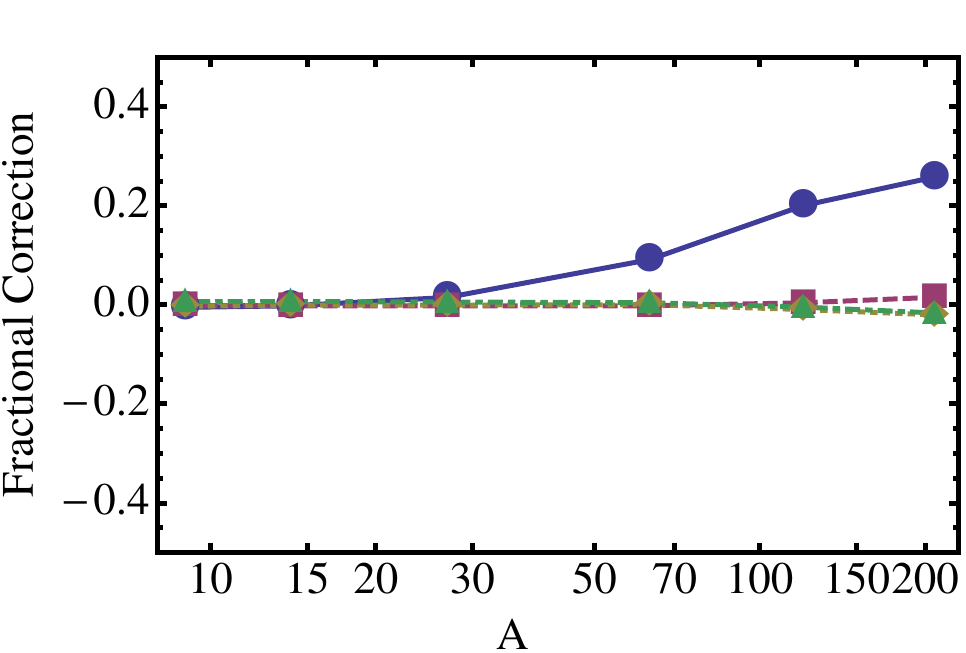}

\caption{Effect of third-order corrections at beam energy of 40 MeV.}
\end{subfigure}
\begin{subfigure}{0.45\textwidth}
\includegraphics[width=\textwidth]{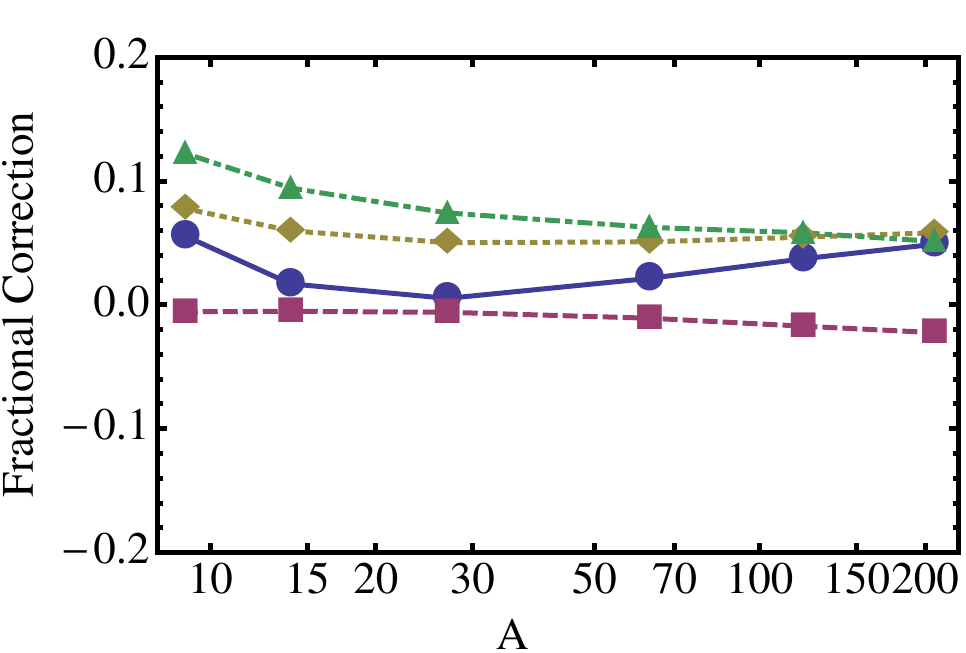}
\caption{Effect of first-order corrections at a beam
  energy of 100 MeV/nucleon at beam energy of 40 MeV.} 
\label{fig:compfrac100MeV}
\end{subfigure}
\caption{(Color online) Fractional corrections at various orders and
  beam energies. The designations for the lines are the same as in
  fig.~\ref{fig:BertschPlot}.}
\label{fig:compfrac}
\end{figure}

In this subsection we present the results of applying the Wallace
corrections to the total cross-sections described above
(eqns.~\ref{Eq:diff},~\ref{Eq:strip}, and~\ref{Eq:absorb}). Our primary
results for these calculations are summarized by
figs.~\ref{fig:BertschPlot}-\ref{fig:compfrac}.

Figure~\ref{fig:BertschPlot} shows the effect of the first order 
corrections for scattering at 40 MeV/nucleon. These corrections are  generally not negligible
for any value of $A$

Figure~\ref{fig:compexp} compares our results to scattering data
collected at 41 MeV/nucleon by Anne et. al.~\cite{Anne:1994}. The data
were collected by detecting the ${}^{10}$Be core, so the processes
that contribute are diffractive scattering, neutron stripping, and
Coulomb breakup, which we did not consider. The Coulomb cross-section
was taken from~\cite{Anne:1994} and added to our
calculations. Although the corrections have a noticeable effect when
compared to the zeroth-order calculations, it is unclear from this
data whether the effect is actually significant since the error in the
measurements is so large. With more precise experimental measurements
the utility of these corrections will become clearer.

Figure~\ref{fig:compfrac} gives the fractional correction at
each order, which more clearly illustrates the effects of the
corrections. The corrections to neutron stripping and total absorption
are only significant at first-order, and appear to be independent of
$A$. The corrections to diffractive scattering are significant at
large values of $A$ all the way through third-order, but are less
significant at low values of $A$. We have also performed the same
calculations at the higher energy of 100 MeV (see
figure~\ref{fig:compfrac100MeV}). As expected, the 
corrections are smaller at first order (less than 10\%), and are less
than 1\% at higher orders.
\begin{figure}[h]
\centering
\includegraphics[width=0.5\textwidth]{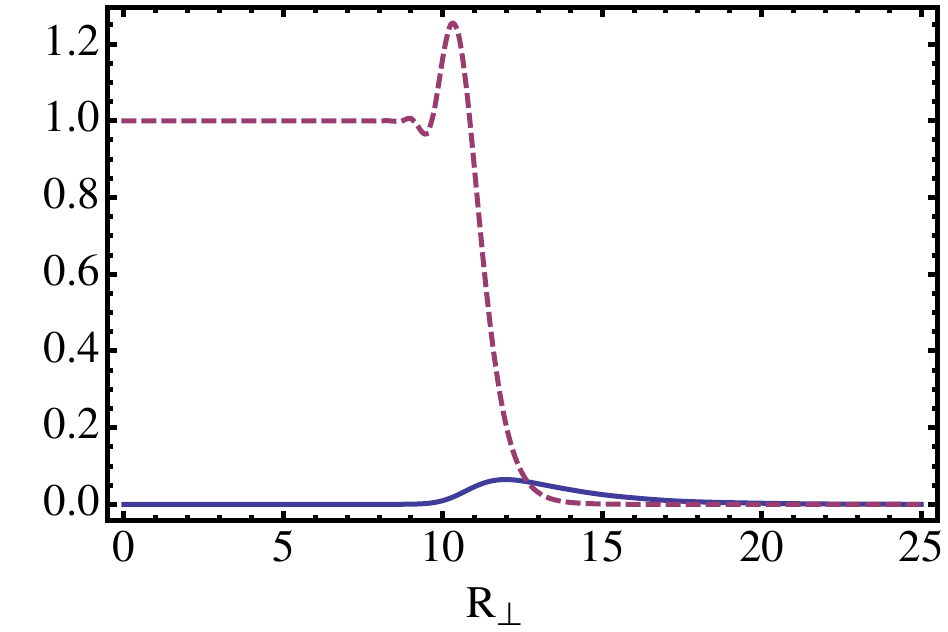}
\caption{(Color online) The integrand of eq.~\ref{Eq:diff} (solid
  blue) for a ${}^{208}$Pb target, and the second term in the same integrand, which is the
  elastic scattering for the system (dashed magenta). Both are given
  as functions of $R_\perp$ (see Fig.~\ref{fig:coords}) with all other
  variables integrated out.}
\label{fig:elvdiffpb}
\end{figure}

Diffractive scattering is primarily a surface effect, (see
Fig.~\ref{fig:elvdiffpb}) which is why
the diffractive corrections have a markedly different behavior as a
function of $A$ than the other types of scattering. As $A$ changes, the radius of
the target nucleus changes as well. The corrections are larger for
surface effects than for volume effects (especially at low energies)
because of the derivative operators $\widehat{\beta}_n$ that arise. Since
diffractive scattering is the only type of scattering studied here
that is almost entirely a surface effect, changes in the radius of the
target affect it more than the other types of scattering we studied.
\section{Summary and Discussion}
\label{sec:summ}
We have calculated corrections to the eikonal approximation to nuclear
scattering in an eikonal expansion framework for many different
processes. We find that for the case of simple potential scattering it
is clear that application of these corrections improves the accuracy
of the eikonal approximation at beam energies between 30 and 100
MeV.  It is reasonable to expect that the first-order correction would
be significant at even higher beam energies.

We also see from application to  the interactions of ${}^{11}$Be with
nuclei at 40 MeV  that these corrections can be as high as 15\% for
neutron stripping and diffractive scattering. As expected, the
corrections decrease as the beam energy increases.

We compare our theory with the data of Anne et al.~\cite{Anne:1994}
and find that the corrections are substantial, although not as large
as the experimental uncertainties.

It is interesting to note that the diffractive corrections have a
strikingly different behavior from the corrections to stripping and
absorption. We attribute this to surface effects that have a stronger
influence on diffractive scattering than on the other processes. 

The first-order corrected cross-sections do not require much more
computational effort to calculate than the zeroth-order
calculations. We performed our calculations on an 8 core node of the
Hyak scientific computing cluster at the University of Washington, and
saw less than a factor of 2 increase in computation time after
including the first-order corrections. Even adding in the second- and
third-order corrections usually resulted in less than a factor of 2
increase in computation time, although the calculation of the
$T$-matrix elements does increase in complexity
(eqns.~\ref{Eq:Tmats}-\ref{Eq:phases}).

Thus we believe that our proposed framework of using the eikonal
approximation as improved by the corrections of Wallace would be a
useful way to analyze data produced at FRIB. Future work will focus on
specific reactions of   experimental interest.


\section*{Acknowledgements} The authors would like to thank George
Bertsch for sharing some of his data,  providing advice on some
calculations and commenting on the manuscript. This work has been partially supported by U.S. D. O. E.
Grant No. DE-FG02-97ER-41014 and by the University of Washington
eScience Institute.
 

\begin{thebibliography}{99}


\bibitem{Wallace:1971zz} 
  S.~J.~Wallace,
  Phys.\ Rev.\ Lett.\  {\bf 27}, 622 (1971).
 
\bibitem{Wallace:1973iu} 
  S.~J.~Wallace,
  Annals Phys.\  {\bf 78}, 190 (1973).
\bibitem{msu}http://www.nscl.msu.edu/future/nsclwhitepaper2006 ``Isotope Science Facility
at Michigan State University
Upgrade of the NSCL rare isotope research capabilities''
MSUCL-1345,
Nov. 2006

\bibitem{Hansen:2003sn} 
  P.~G.~Hansen and J.~A.~Tostevin,
  Ann.\ Rev.\ Nucl.\ Part.\ Sci.\  {\bf 53}, 219 (2003).

\bibitem{gg08} A.~Gade and T.~Glasmacher, 
Prog. Part. Nucl. Phys. 60, 161 (2008)


\bibitem{Bertulani:2009mf}    C.~A.~Bertulani and A.~Gade,
  Phys.\ Rept.\  {\bf 485}, 195 (2010)
 
 \bibitem{roy}R.~J.~Glauber, ``High Energy Collision Theory",p. 315  in ``Lectures in Theoretical Physics" Ed. by 
W.~E.~Brittain and L.~G.~Dunham, Vol. I, Interscience, New York, 1959
 R.~J.~Glauber,
  ``Theory of high energy hadron-nucleus collisions,''
  p. 207,  In  ``High-Energy Physics And Nuclear Structure", ed. by S. Devons, Plenum Press,  New York 1970



\bibitem{Bertsch:1990zz} 
  G.~Bertsch, H.~Esbensen and A.~Sustich,
  Phys.\ Rev.\ C {\bf 42}, 758 (1990).
 
\bibitem{Ogawa:1992tf} 
  Y.~Ogawa, K.~Yabana and Y.~Suzuki,
  Nucl.\ Phys.\ A {\bf 543}, 722 (1992).

\bibitem{AlKhalili:1996zz} 
  J.~S.~Al-Khalili, J.~A.~Tostevin and I.~J.~Thompson,
  Phys.\ Rev.\ C {\bf 54}, 1843 (1996).

 
\bibitem{Aumann:2000zz} 
  T.~Aumann, A.~Navin, D.~P.~Balamuth, D.~Bazin, B.~Blank, B.~A.~Brown, J.~E.~Bush and J.~A.~Caggiano {\it et al.},
  Phys.\ Rev.\ Lett.\  {\bf 84}, 35 (2000).

\bibitem{Hencken:1996af} 
  K.~Hencken, G.~Bertsch and H.~Esbensen,
  Phys.\ Rev.\ C {\bf 54}, 3043 (1996)

\bibitem{Parfenova:2000be} 
  Y.~. L.~Parfenova, M.~V.~Zhukov and J.~S.~Vaagen,
  Phys.\ Rev.\ C {\bf 62}, 044602 (2000).

\bibitem{Licot:1997zz} 
  I.~Licot, N.~Added, N.~Carlin, G.~M.~Crawley, S.~Danczyk, J.~Finck, D.~Hirata and H.~Laurent {\it et al.},
  Phys.\ Rev.\ C {\bf 56}, 250 (1997).


\bibitem{Sauvan:2000hq} 
  E.~Sauvan, F.~Carstoiu, N.~A.~Orr, J.~C.~Angelique, W.~N.~Catford, N.~M.~Clarke, M.~Mac Cormick and N.~Curtis {\it et al.},
  Phys.\ Lett.\ B {\bf 491}, 1 (2000)


\bibitem{Bertulani:2006fe} 
  C.~A.~Bertulani and A.~Gade,
  Comput.\ Phys.\ Commun.\  {\bf 175}, 372 (2006)
%
\bibitem{Esbensen:2001mr} 
  H.~Esbensen and G.~F.~Bertsch,
  Phys.\ Rev.\ C {\bf 64}, 014608 (2001).
\bibitem{Wallace:1973ni} 
  S.~J.~Wallace,
  Phys.\ Rev.\ D {\bf 8}, 1846 (1973).


\bibitem{Varner:1991zz} 
  R.~L.~Varner, W.~J.~Thompson, T.~L.~McAbee, E.~J.~Ludwig and T.~B.~Clegg,
  Phys.\ Rept.\  {\bf 201}, 57 (1991).

\bibitem{devries87} H.~de~Vries, C.~W.~de~Jager, and C.~de~Vries,
  At. Data Nucl. Data Tables {\bf36}, 495 (1987).

\bibitem{Anne:1994}
  R.~Anne,  {\it et al.},
  Nuc. Phys. {\bf A575}, 125 (1994).

 \end{thebibliography}
\end{document}